\documentclass[prd,superscriptaddress]{revtex4}
\usepackage{graphicx}
\usepackage{epsf}
\usepackage{epsfig}
\usepackage{amssymb}
\usepackage{amsmath}

\def\1{{\'{\i}}}

\newcommand{\df}{\displaystyle\frac}
\newcommand{\be}{\begin{equation}}
\newcommand{\ee}{\end{equation}}
\newcommand{\bea}{\begin{eqnarray*}}
\newcommand{\eea}{\end{eqnarray*}}

\begin{document}
\title{The Integrated Sachs-Wolfe Effect in Interacting Dark Energy Models}
\author{Germ\'{an} Olivares\footnote{E-mail address: german.olivares@uab.es}}
\affiliation{Departamento de F\'{\i}sica, Universidad Aut\'{o}noma de Barcelona,
Barcelona, Spain}
\author{Fernando Atrio-Barandela\footnote{E-mail address: atrio@usal.es}}
\affiliation{Departamento de F\'{\i}sica Fundamental, Universidad
de Salamanca, Spain}
\author{Diego Pav\'{o}n\footnote{E-mail address: diego.pavon@uab.es}}
\affiliation{Departamento de F\'{\i}sica, Universidad Aut\'{o}noma de Barcelona,
Barcelona, Spain}

\begin{abstract}
Models with dark energy decaying into dark matter have been
proposed in Cosmology to solve the coincidence problem. We study
the effect of such coupling on the cosmic microwave background
temperature anisotropies. The interaction changes the rate of
evolution of the metric potentials, the growth rate of matter
density perturbations and modifies the Integrated Sachs-Wolfe
component of cosmic microwave background temperature anisotropies,
enhancing the effect. Cross-correlation of galaxy catalogs with
CMB maps  provides a model independent test to constrain the
interaction. We particularize  our analysis for a specific
interacting model and show that galaxy catalogs with median
redshifts $z_m=0.1-0.9$ can rule out models with an interaction
parameter strength of $c^2\simeq 0.1$ better than 99.95\%
confidence level. Values of $c^2\le 0.01$ are compatible with the
data and may account for the possible discrepancy between the
fraction of dark energy derived from WMAP 3yr data and the
fraction obtained from the ISW effect. Measuring the fraction of
dark energy by these two methods could provide evidence of an
interaction.
\end{abstract}
\pacs{98.80.Es, 98.80.Bp, 98.80.Jk}
\maketitle

\section{Introduction}
Measurements of luminosity distances using supernovae type Ia
(SNIa) \cite{riess}, of the cosmic microwave background (CMB)
temperature anisotropies with the WMAP satellite \cite{wmap3},
large scale structure \cite{lss,cole}, the integrated Sachs--Wolfe
effect \cite{fosalba-gazta2003,boughn-crittenden},
and weak lensing \cite{weakl}, indicate
that the Universe has entered a phase of accelerated expansion
-see \cite{reviews} for recent reviews.
This acceleration is explained in terms of an unknown and
nearly unclustered matter component of negative pressure,
dubbed ``dark energy" (energy density,
$\rho_{x}$), that currently contributes with about $75\%$ of the total
density. The remaining $25\%$ is shared between cold dark matter
($\rho_{c} \sim 21\%$), baryons ($\rho_{b} \sim 4\%$), and
negligible amounts of relativistic particles (photons and neutrinos).
The pressureless non-relativistic matter component redshifts
faster with expansion than the dark energy, giving rise to the so-called
`coincidence problem' that seriously affects many models
of late acceleration, particularly the $\Lambda$CDM model:
``Why are matter and dark energy densities
of the same order precisely today?" \cite{steinhardt}.
To address this problem within general relativity one must either
accept an evolving dark energy field or adopt an incredibly tiny
cosmological constant and admit that the ``coincidence" is just
that, a coincidence that might be alleviated by turning to
anthropic ideas \cite{anthropic}. (It should be mentioned,
however, the existence of proposals in which a vacuum energy
density of about the right order of magnitude arises from the
Casimir effect at cosmic scales -see \cite{emili} and references
therein).

One way to address the coincidence problem, within general
relativity, is to assume an interaction (coupling) between the
dark energy component and cold dark matter such that the ratio $r
\equiv \rho_{c}/\rho_{x}$ evolves from a constant but unstable
value at early times (in the radiation and matter dominated
epochs) to a lower, constant and stable attractor at late times
well in the accelerated expansion era \cite{iqm0,iqm}. Before the
late accelerated expansion was observed, Wetterich introduced
interacting quintessence models to reduce the theoretical high
value of the cosmological constant \cite{wetterich}. Later, these
kind of models were rediscovered, sometimes (but not always) in
connection with the coincidence problem -see, e.g.
\cite{in-connection}. Other solutions (known as $f(R)$
models) require the Einstein-Hilbert action to be modified \cite{f(R)}.

Interacting quintessence models are testable since they predict
differences in the  expansion rates of the Universe, the growth of
matter density perturbations, the Cosmic Microwave Background
(CMB) temperature anisotropies and in other observables. In this
paper we shall demonstrate that, as the Integrated Sachs-Wolfe effect
measures directly the growth rate of matter density perturbations,
can be used to detect variations on the evolution of
the large scale structure with respect to the prediction of the concordance $\Lambda$CDM
model. As a toy model,  we shall consider a spatially flat
Friedmann--Robertson--Walker universe filled with radiation,
baryons, dark matter and dark energy such that the last two
components interact with each other through a coupling term, $Q =
3 {\cal H} c^{2} (\rho_{c} +\rho_{x})$. Thus, the energy balance
equations for dark matter and dark energy take the form
\\
\begin{equation}
\dot{\rho}_{c}+3{\cal H} \rho_{c} = 3 {\cal H} c^{2}
(\rho_{c} +\rho_{x})\,  ,\qquad
\dot{\rho}_{x}+3 {\cal H} (1+w_{x})\rho_x= -3{\cal H} c^{2}
(\rho_{c} +\rho_{x}) \, , \label{balance}
\end{equation}
\\
where $w_{x} = p_{x}/\rho_{x} < -1/3$ is the equation of state
parameter of the dark energy component and $c^{2}$ is a constant
parameter that measures the strength of the interaction (it should
not be confused with the speed of light that we set equal to
unity). Derivatives are taken with respect to the conformal time,
and ${\cal  H}= \dot a/a$, with $a$ the scale factor of the
Friedmann-Robertson-Walker metric. To satisfy the severe
constraints imposed by local gravity experiments
\cite{peebles_rmph, hagiwara}, baryons and radiation couple to the
other two energy components only through gravity. Our ansatz for
$Q$ guarantees that the ratio between energy densities, $r$, tends
to a fixed, attractor, value at late times. It yields  a constant
but unstable ratio at early times. The details of the calculation
can be found in \cite{iqm,olivares1,olivares2,olivares3}. This
result holds irrespective of whether the dark energy is a
quintessence (i.e., $ -1 <w_{x} < -1/3$), phantom ($w_{x} < -1$),
$k$-essence or tachyon scalar field \cite{dualk}. It is worth to
remark that the above ansatz implies  an effective power law
potential for the quintessence dark energy field at early times
and an effective exponential potential at late times \cite{iqm0}.

Beyond the physical motivation for the existence of a coupling
between dark matter and dark energy, their interaction parameter
$c^{2}$ needs to be constrained by observations. In
\cite{olivares2} we showed that the slope of the matter power
spectrum at small scales decreases and the scale of matter
radiation equality is shifted to larger scales with the
interaction.  CMB temperature anisotropies are equally affected if
the interaction is present at decoupling. The luminosity distance
of SNIa depends on the expansion rate of the Universe and can be
used to test the existence of a DM/DE coupling. Comparison of
model predictions with data showed that only WMAP 3yr data and the
Sloan Digital Sky Survey (SDSS) data provided constraints on the
value of $c^2$ \cite{olivares2,olivares3}. In particular, the SNIa
data proved to be rather insensitive to the coupling. In this
paper we show that the Integrated Sachs-Wolfe (ISW) component of
the CMB temperature anisotropies is much more sensitive than the
SNIa data at redhisfts $z\sim 1$. It measures the growth factor of
linear matter density perturbations, that is strongly affected by
the interaction and can effectively set constraints on Interacting
Quintessence Models (IQM). Other experiments that also constrain
the growth rate are discussed in \cite{amendola}. In Section II we
briefly summarize the formalism behind the measurements of the ISW
effect by cross-correlating CMB data with galaxy catalogs. In
Section III we describe the observational prospects and in Section
IV we present our conclusions.

\section{The Integrated Sachs-Wolfe effect}
When CMB photons in their way to the observer from the surface
of last scattering, cross a time-varying potential well, experience
the so called late Integrated Sachs-Wolfe effect. Photons are
blueshifted (redshifted) when entering (leaving) a high dense region,
and similarly for low dense regions.
If the gravitational potential $\phi$ evolves during a photon
crossing like in models with a cosmological constant, then both
effects will not compensate each other and the final energy of the
photon will vary. The net effect will be \cite{sachs-wolfe},
\\
\begin{equation}
\frac{\Delta T_{ISW}(\hat{n})}{T_{0}} =-2\int_0^{r_{rec}} dr \,
\frac{\partial\phi(r,\hat{n})}{\partial r}. \label{eq:isw}
\end{equation}
\\
The integral is taken with respect to conformal distance (or
look-back time) from the observer at redshift $z=0$, $r(z)=
\int_0^{z} dz'/H(z')$ where the Hubble factor, $H$, obeys
Friedmann's equation, $3 H^{2} = 8\pi G (\rho_{c}+\rho_{x})$.

\begin{figure}[t]
\epsfxsize=0.85\columnwidth\epsfbox{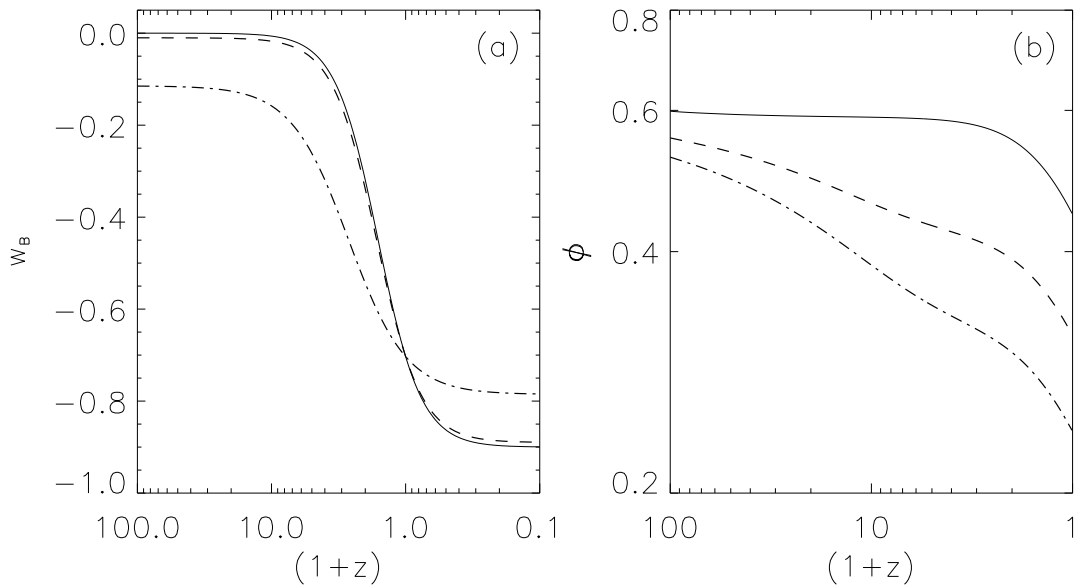} \caption{Panel
(a): Evolution of the background equation of state parameter, $w_{B}$, as a function
of redshift. Panel (b): Evolution of the Newtonian potential
$\phi$ for the wavenumber mode $k=0.01h\, $Mpc$^{-1}$. Solid,
dashed and dot-dashed lines correspond to $c^2=0,0.01$, and $0.1$,
respectively. In drawing the figures we have set $w_{x} = -0.9$. }
\label{fig1}
\end{figure}

The evolution of the Newtonian potential will be computed
numerically, but its qualitative behavior can be understood from
the perturbed Einstein's equations. Once the Universe becomes
matter dominated, we can write in the conformal Newtonian gauge
\cite{olivares2,ma}:
\\
\begin{equation}
\ddot\phi +{\cal H}\dot\phi -3w_B{\cal H}^2\phi = \frac{3}{2}{\cal H}^2\sum_{i}
\left(\frac{\delta P_i}{\delta\rho_i}\right)^{2}\delta_i\Omega_i \, ,
\label{hgravitation}
\end{equation}
\\
where the sum is over all matter fluids and scalar fields,
$\Omega_i$, $\delta_i$ are the  density parameter and the energy
density contrast, respectively, of fluid $i$. Derivatives are also
taken with respect to conformal time. In non-interacting models,
the gravitational potential changes when (a) the equation of the
state of the background varies, and (b) the dark energy is
smoothly distributed, i.e., it does not cluster on scales below
the ``sound horizon", set by $c_{s,x}^2=(\delta
P_x/\delta\rho_x)$. At that point the potential will decay and
generate an ISW effect \cite{hu98, hu-scranton}. During the matter
dominated epoch, the matter sound speed is $c_{s,c}^2=(\delta
P_c/\delta\rho_c)=0$. If $c_{s,x}^{2}=1$, then the dark energy
fluid  does not cluster on scales below the horizon
\cite{erickson}, whereby $\delta_x \sim 0$. In this case, on
subhorizon scales the gravitational potential evolves as
$\phi=C_1+C_2a^{-5/2}$, i.e., it essentially remains constant.
However, in models with interaction, the CDM energy density does
not scale as $a^{-3}$ because of the coupling. During the matter
dominated regime, the background equation of state parameter is
$w_B=c^2w_x/(c^2+1)<0$ and the growing mode in Eq.
(\ref{hgravitation}) becomes $\phi= C_{1}\, a^m$ with
$m<3w_B/4<0$; that is, there are no growing modes and the
potential $\phi$ decays. In the period of accelerated expansion,
in both interacting and non-interacting models, $\phi$ decays.
In Fig.~\ref{fig1}a we plot the evolution of $w_B$ as a function
of redshift. Solid, dashed and dot-dashed lines correspond to
$c^2=0, 0.01$, and $0.1$, respectively, and we have set
$w_x=-0.9$. Notice that when $c^2\ne 0$, the matter dominated
period has $w_B\ne 0$. In Fig.~\ref{fig1}b we depict the time
evolution of the Newtonian potential $\phi$ as a function of
redshift for  a mode of wavenumber $k=0.01h\,$Mpc$^{-1}$. In the
three cases, the amplitude of the  potential is normalized to
unity at horizon crossing. We solved the system of equations that
describes the evolution of all energy components and the Newtonian
potentials  using a modified version of the CMBFAST code
\cite{seljak}. As the figure shows, when $c^2=0$, $\phi$ remains
constant during the matter dominated epoch, and it decays solely
during the phase of accelerated expansion. By contrast, in
interacting models  the potential $\phi$ decays even during matter
domination.

The ISW effect is not the only anisotropy generated by local structures.
Secondary anisotropies could be generated due to non-linear evolution of the
gravitational potential, usually associated with the gravitational collapse
of small scale structures like clusters \cite{rees-sciama}.
But the relevant scales are those of clusters and superclusters, corresponding to
angular scales of $5-10$ arcmin, much smaller than those associated to the ISW effect.

\begin{figure}[t]
\epsfxsize=\columnwidth\epsfbox{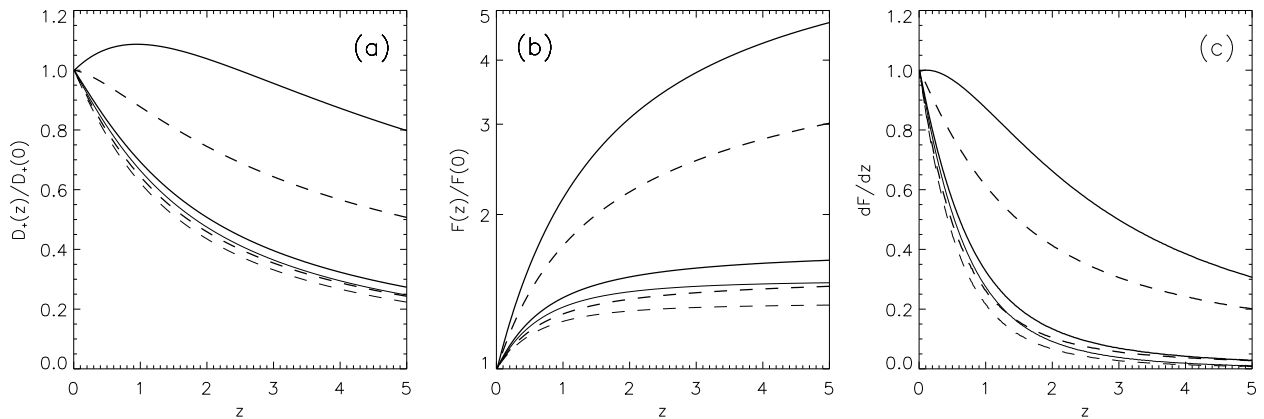} \caption{(a) growth
rate of matter density perturbations, (b) growth rate relative to Einstein-de Sitter and (c)
differential growth rate per unit of redshift for two cosmological
models with $w_x=-0.9$ and different dark energy fraction and
interaction parameter. In all panels, solid lines correspond to
$\Omega_{x}=0.8$, and dashed lines to $\Omega_{x}=0.7$. In each
set and from top to bottom, the interaction parameter is $c^2=
0.1$, $0.01$, and $0$. } \label{fig:growth}
\end{figure}

The ISW component cannot be measured directly, i.e., cannot be extracted
from CMB data using frequency information since it
the same frequency dependence that any other intrinsic CMB
signal. Since at low redshifts is generated by
local structures, the spatial pattern of ISW
anisotropies will correlate with the matter distribution in the local Universe.
The contribution to the radiation power spectrum due to the evolution
of the gravitational potential at low redshifts is given by \cite{crittenden-turok,cooray}
\\
\begin{equation}
C_l^{ISW}=<|a_{lm}|^2>=\frac{2}{\pi}\int_0^\infty k^{2} \, dk \,
P(k)\, I_l^2(k) \, , \label{eq:cl_isw}
\end{equation}
\\
where
\\
\begin{equation}
I_l(k)=3\Omega_{c,0}\, \frac{H_{0}^{2}}{
k^{2}}\int_0^{r_{rec}}dr\, j_l(kr)\, \frac{dF}{dr} \, .
\label{eq:Il_isw}
\end{equation}
\\
In this expression, $F=D_+(z)(1+z)$ is the ratio of the growth of
matter density perturbations in any cosmological model, $D_+(z)$,
to that in a Einstein-de Sitter Universe, $(1+z)^{-1}$;
$H_{0}^{-1}$ is the present Hubble radius and $j_l(x)$ are
spherical Bessel functions. In Figure \ref{fig:growth} we plot the
redshift evolution of $D_+$, $F$ and $dF/dz$. All the models have
$w_x=-0.9$ and solid lines correspond to $\Omega_{x} =0.8$ and
dashed lines to $\Omega_x =0.7$. From top to bottom, each group of
lines corresponds to $c^2=0.1,0.01$, and $0$, the latter
pertaining to non-interacting models. The growth function was
normalized to unity at $z=0$,  $D_+(0)=1$.

Due to the ISW component, CMB temperature maps will correlate with
tracers of the gravitational potential at low redshifts that could
be constructed from galaxy catalogs. In particular, it will
correlate with templates of the projected galaxy density along the
line of sight. Assuming the distribution of galaxies follows that
of the dark matter, the correlation of a template of projected
galaxy density and a CMB map can be computed analytically
\cite{crittenden-turok},
\\
\begin{equation}
C_{l}^{ISW-M}=\df{2}{\pi}\int k^2 dk \, P(k) \, I_l^{ISW}(k) \,
I_{l}^{M}(k) \, , \label{eq:cl_iswM}
\end{equation}
\\
with
\\
\begin{equation}\label{window_T}
I_l^M=\int_0^{r_0} dr \, W^M(k,r) \, j_l(kr)\, ,
\end{equation}
\\
where $W^{M}(k,r) = b(k,r) \, n_M(r)\, D_+(r)$ is the window
function of the tracer of the large scale gravitational potential,
$b(k,r)$ is the scale dependent bias existing between sources and the matter distribution
as a function of radial
distance $r$, $n_M(r)$ is the normalized redshift distribution of
sources and $D_+(r)$ is the growth function of matter density perturbations.
Similarly, the power spectrum of the galaxy catalog
is given by
\\
\begin{equation}
C_l^{M}=(2/\pi)\int k^2P(k)(I_l^M(k))^2dk.
\end{equation}
\\
One  advantage of using Eq. (\ref{eq:cl_iswM}) is that it
separates the contribution coming from redshift evolution and from
the spatial distribution of the large scale structure. Models of
structure formation that give rise to the same matter power
spectrum but differ in their growth rate, $D_+(z)$ (see
Fig.~\ref{fig:growth}b), will also give rise to ISW effects of
different amplitude.  Consequently, the correlation with galaxy
catalogs will also differ. If the interaction varies with time and
is only significant at present, the shape of the matter power
spectrum will be unaffected but the growth rate will not.
Therefore, the ISW effect will be sensitive to the existence of
the interaction  through the influence of the latter on the growth
rate of density perturbations. In the next section, we shall fix
the matter power spectrum of all models with interaction to be the
same as a model with $c^2=0$ (no interaction) and the same
cosmological parameters. We considered only power spectra that
were compatible in amplitude and shape with the spectra derived
from the SDSS and 2dGF galaxy catalogs \cite{cole}. Thus,
differences in the correlation between templates of projected
galaxy density and CMB data -Eq.(\ref{eq:cl_iswM})- will reflect
differences in the growth ratio $F$ only and will indicate how
sensitive the ISW effect is to the redshift evolution of matter
density perturbations.

\section{Observational prospects}
After the release of WMAP 1yr data, several groups looked for
evidence of ISW effect by cross-correlating the WMAP data with
templates built from different catalogs: the NVSS radio survey
\cite{nolta,vielva-mcewen}, the HEA0-I X-ray Background map
\cite{boughn-crittenden}, the APM \cite{fosalba-gazta2004}, the
SDSS \cite{fosalba-gazta2003,padmanabhan,cabre}, and the 2MASS
galaxy surveys \cite{afshordi-2mass,rassat}. These analyses rely
on constructing templates from catalogs as tracers of the ISW
signal. Cross-correlation of random realizations of CMB data and
noise provides an estimate of the uncertainties. Catalogs
differed in depth and sky coverage and different groups use
different techniques: real space correlation, correlations in
Fourier space, wavelets analysis, etc., and obtained positive
detections with levels of significance of $2-4\sigma$. These
detections represent an independent evidence for the existence of
dark energy, though, at the moment, the measurements are not
accurate enough to discriminate between competing dark energy
scenarios.

In this section, we analyze the prospects that the ISW effect
could be potentially sensitive to the growth rate of matter
density perturbations and, therefore, to the dark energy--dark
matter coupling.  The effect of the interaction is to increase the
amplitude of the cross-correlation of template and CMB data: the
interaction makes the gravitational potentials  decay faster which
produces a larger ISW effect, making it easier to detect. If we
trace the ISW effect with a template $M$ that covers a fraction
$f_{sky}$ of the sky, the cumulative signal to noise ratio is
given by \cite{cooray}
\\
\begin{equation}
\left(\frac{S}{N}\right)^2=f_{sky}\sum_{l=2}^{l_{max}}(2l+1)
\frac{(C_l^{ISW-M})^2}{(C_l^{ISW-M})^2+(C_l^{CMB}C_l^M)} \, ,
\label{eq:sn}
\end{equation}
\\
where $C_l^M$ is the power spectrum of the template $M$. The sum
extends to the largest angular scale resolved by the template. The
main source of confusion in ISW measurements comes from the
intrinsic CMB anisotropies. In Eq. (\ref{eq:sn}) we have neglected
the instrumental noise of WMAP data and the shot noise of the
template map.

\begin{figure}[t]
\epsfxsize=0.6\columnwidth\epsfbox{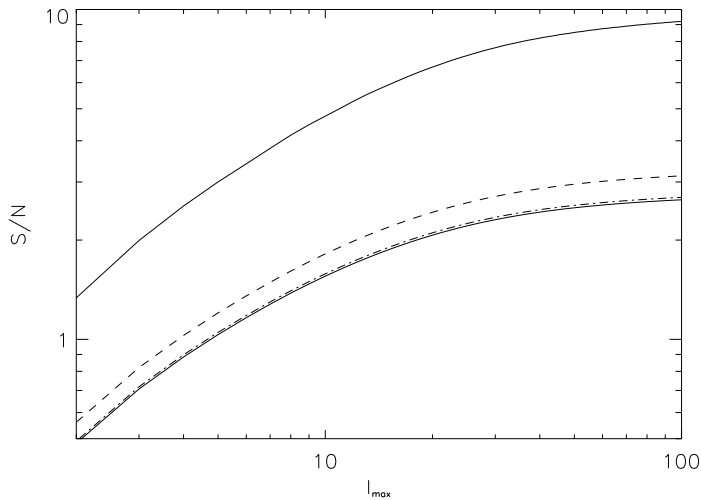}
\caption{Cumulative signal to noise ratio for a catalog with mean
depth $z_m=0.3$.  From bottom to top lines correspond to $c^2=0$
(bottom thin-solid line), $10^{-3}$ (dot dashed line), $0.01$
(dashed line), and $0.1$ (solid line). Template and data are
assumed to have full sky coverage.} \label{fig:signal_noise}
\end{figure}

If the distribution of galaxies as a function of redshift is
known, then the expected S/N ratio can be computed numerically.
Following earlier practice, we model the redshift distribution of
sources by the analytic expression,
$n_M(z)=A(z/z_{0})^2\exp[-(z/z_{0})^{3/2}]$. The constant $z_{0}$
measures the effective depth of the catalog and is simply related
to the median redshift $z_m$ of the source distribution,
$z_m=1.412\, z_0$ \cite{cabre2007,gazta2006}. The normalization
constant is fixed by setting $\int n_M(z) dz=1$. In
Fig.~\ref{fig:signal_noise} we plot the cumulative S/N ratio for a
cross-correlation of temperature data with a template constructed
from a galaxy catalog with median redshift $z_m=0.3$, as a
function of the multipole $l_{max}$ for 4 different interaction
parameters. From bottom to top $c^2=0$ (bottom thin-solid line),
$10^{-3}$ (dot dashed line, almost coincident with the previous
one), $0.01$ (dashed line), and $0.1$ (solid line). For simplicity
we considered data and template to have full sky coverage, i.e.,
$f_{sky}=1$. In Fig.~\ref{fig:signal_noise} it is seen that
increasing the strength of the interaction augments the S/N ratio
reflecting the fact that, as the interaction generates larger ISW
signal, it can be detected with higher statistical significance.

\begin{figure}[t]
\epsfxsize=0.6\columnwidth\epsfbox{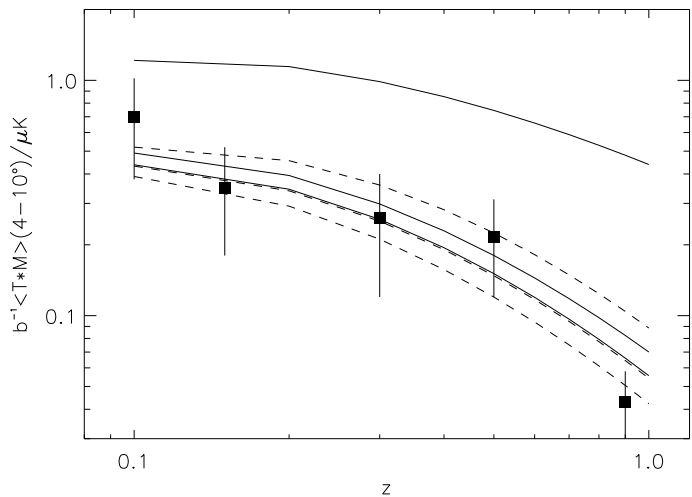}
\caption{ Cross-correlation between CMB
data and template, averaged on angular scales from $4$ to $10^o$.
Dashed lines corresponds to models with no interaction ($c^{2} =
0$) and $w_x=-0.9$ and, from bottom to top: $\Omega_{x}=0.7,
0.74$, and  $0.8$. Solid lines correspond to interacting models
(from bottom to top: $c^2=10^{-3}, 0.01$, and $0.1$), $w_{x} =
-0.9$, and $\Omega_{x}=0.74$. Full diamonds represent the
measurements obtained from different catalogs and their $1\sigma$
errors compiled in \cite{gazta2006}. } \label{fig4}
\end{figure}

Comparison between model predictions and observations can be done
by means of the correlation function. The theoretical prediction
is $<T*M>(\theta)=\sum_l(2l+1)C_l^{ISW-M}P_l(\cos\theta)/4\pi$.
The cross-correlation of a template M constructed from a tracer of
the large scale matter distribution like a galaxy catalog with a
CMB map is simply
\\
\begin{equation}
\langle T*M\rangle (\theta) =  \frac{\sum_{(i,j)\in\theta} M_i
T_j/\sigma^2_j} {\sum_{(i,j)\in\theta} 1/\sigma^2_j} \, ,
\label{eq:isw-correlation}
\end{equation}
\\
where $M_i$ is the projected galaxy density on pixel $i$ and
$\sigma_i$ is the noise associated to each pixel on a CMB map. The
measured correlation function in Eq. (\ref{eq:isw-correlation}) is
affected by masking, pixelization and the inhomogeneities within
the catalog of the galaxy selection function across the sky.
Further, the free electrons that reside in the deep potential
wells of galaxy clusters contribute to the temperature
anisotropies via the thermal Sunyaev-Zeldovich (SZ) effect
\cite{atrio-muc}, that dilute the ISW signature at zero lag
\cite{fosalba-gazta2003}. For example, templates constructed with
the 2MASS catalog of galaxies are known to trace the thermal SZ
component \cite{monteagudo} as well the ISW effect. To avoid a
possible thermal SZ contamination, we will compare the mean
between $4$ and $10^o$ of the predicted and measured correlation
functions. In Figure \ref{fig4} we plot the correlation function
computed using the analytic galaxy redshift distribution given
above. Theoretical estimates were done assuming the bias factor
$b$ to be constant. Dashed lines corresponds to models with no
interaction ($c^2=0), w_x=-0.9$ and three different dark energy
densities: from bottom to top $\Omega_x=0.7,0.74$, and $0.8$.
Solid lines correspond to interacting models with $c^2=10^{-3},
0.01$, and $0.1$ (from bottom to top), $w_x=-0.9$,  and
$\Omega_x=0.74$. Full diamonds represent measurements of the ISW
amplitude obtained by cross-correlating different galaxy catalogs
with WMAP first year data (see \cite{gazta2006} for details). For
consistency with the data, the theoretical estimates are given in
units of $b$. In the figure, the matter power spectra was
normalized fixing the amplitude of the rms of matter density
perturbations on a sphere of $8h^{-1}$Mpc to be $\sigma_8=0.8$. As
expected, if all models have the same $\sigma_8$, the ISW effect
increases with increasing $\Omega_x$ and increasing $c^2$. The
$\chi^2$ per degree of freedom is less than $1$, except for
$c^2=0.01$, $\Omega_x=0.74$ ($\chi^2_{dof}=1.5$) and $c^2=0$,
$\Omega_x=0.8$ ($\chi^2_{dof}=3.6$). An interaction parameter as
high as $c^2=0.1$ is strongly ruled out by the observations,
$\chi^2_{dof}=180$. The low quality of the fit is dominated by the
measurement at $z=0.9$ obtained in \cite{boughn-crittenden} and
derived using NVSS and HEAO-1 samples. Due to the systematic
uncertainties of the  bias and selection function in both samples,
our simple model may not be accurate enough to make a meaningfull
prediction; but even if we disregard this data point, a value of
$c^2=0.1$ can be ruled out at the 99.95\% confidence level. The
rejection level would be smaller if we considered the uncertainty
in the shape of $n_M(z)$ and in the mean depth of galaxy catalogs.
At any rate, this analysis shows that the ISW effect can
effectively constrain the existence of an interaction at low
redshift. It has more statistical power than the luminosity
distance test based on SNIa observations \cite{olivares3}.

If WMAP first year data showed a good agreement between the best
fit values of $\Omega_{x}$ derived from the analysis of
temperature anisotropy maps and those obtained by
cross-correlating temperature maps with projected galaxy density
templates, Cabr\'e {\em et al.} \cite{cabre,cabre2007} noticed
some tension between the values derived from these techniques
using the less noisy WMAP third year data. The ISW signature of
the new CMB data had a preference for larger values of $\Omega_x$.
The combined analysis of different samples gave
$\Omega_{x}=0.77-0.86$ at the 95\% confidence level.  The data was
compatible with $w_{x}=-1$ ($\Lambda$CDM model), but did not have
enough statistical power to break the degeneracy in the $w_{x}$ vs
$\Omega_{x}$ plane. Those results were later confirmed in
\cite{rassat} where a best fit value of $\Omega_\Lambda = 0.85$
for a flat $\Lambda$CDM model was reported, albeit with larger
uncertainty. An interaction parameter of $c^2\sim 0.01$ could
produce a ISW signal similar to a non-interacting model with
higher $\Omega_x$, bringing back the agreement between the
measured fraction of dark energy with WMAP 3yr data and with the
ISW effect. Should the Cabr\'e {\em et al.} or Rassat {\em et al.}
results \cite{cabre,cabre2007,rassat} be confirmed by  deeper
catalogs like the new SDSS releases or PanSTARRS \cite{pan}, the
ISW test will constitute an indirect evidence that the growth rate
of matter density perturbations differs from that of the
concordance $\Lambda$CDM model. Future surveys will be able to
test different extensions of the standard model and prove/rule out
the existence of an interaction.

\section{Conclusions}
The ISW effect may provide an excellent tool to detect, in a model
independent way, the coupling between dark energy and dark matter
at low redshifts, i.e., $z\le 1$. The interaction damps the growing mode
of the Newtonian potential faster compared to models with no
interaction and  enhances the ISW effect; then, the ISW effect is
sensitive to the growth rate of matter density perturbations, but
model predictions need to be refined to model more realistically
the selection function of galaxy catalogs. We have shown that the
correlation of WMAP with templates constructed from different
galaxy catalogs can rule out an interaction parameter of $c^2=0.1$
at better than the 99.95\% confidence level. For this type of models, this
upper limit is stronger than those obtained previously using the
CMB radiation power spectrum \cite{olivares1,olivares3}, but at
a different redshift. The ISW effect shows more statistical
power to discriminate among IQM models than the luminosity
distance test obtained using SNIa.

The recent results in \cite{cabre,cabre2007} suggest a discrepancy
between the fraction of dark energy necessary to explain the
amplitude of the ISW effect (measured from cross-correlating CMB
data and galaxy catalogs) larger than allowed by the WMAP 3yr
data. We have shown that the tension between both data sets could
be explained by introducing a moderate coupling between dark
energy and dark matter. This is in line with recent claims that
the Layzer-Irvine equation \cite{layzer} when applied to  galaxy
clusters reveals the existence of a DM/DE coupling
\cite{clusters}. If the aforesaid discrepancy is confirmed by
deeper and wider galaxy samples, it could turn out an indirect
evidence for the existence of a DM/DE interaction.

\acknowledgments {We are grateful to Wayne Hu for indicating us
Refs. \cite{hu-scranton} and \cite{hu98} as well as useful
correspondence. This research was partly supported by the Spanish
``Ministerio de Educaci\'on y Ciencia" (Grants
FIS2006-12296-C02-01, BFM2000-1322 and PR2005-0359), the ``Junta
de Castilla y Le\'{o}n" (Project SA010C05), and the ``Direcci\'{o}
General de Recerca de Catalunya" (Grant 2005 SGR 00087).}

\end{document}